\documentstyle[sprocl]{article}
\input {epsf}

\newdimen\hhsize\hhsize=.5\hsize

\def\be{\begin{equation}} \def\ee{\end{equation}}
\def\bea{\begin{eqnarray}} \def\eea{\end{eqnarray}}
 \def\be{\begin{equation}}
\def\ee{\end{equation}} \def\bea{\begin{eqnarray}}
\def\eea{\end{eqnarray}} 

\def\lta{\mathrel{\mathpalette\fun <}}
\def\gta{\mathrel{\mathpalette\fun >}}

\def\cmm2{{\,\rm cm^{-2}}}
\def\cm2{{\,{\rm cm}^2}}
\def\cmm3{{\,{\rm cm}^{-3}}}
\def\gcmm3{{\,{\rm g\,cm^{-3}}}}

\def\fun#1#2{\lower3.6pt\vbox{\baselineskip0pt\lineskip.9pt
  \ialign{$\mathsurround=0pt#1\hfil##\hfil$\crcr#2\crcr\sim\crcr}}}

\begin{document}
\title{
Constraints on Compact Hyperbolic Spaces from COBE
}
\author{J. Richard Bond, Dmitry Pogosyan \& Tarun Souradeep }
\address{CITA, University of Toronto, Toronto, ON M5S 3H8, CANADA}

\maketitle\abstracts{The (large angle) COBE DMR data can be used to
probe the global topology of our universe on scales comparable to and
just beyond the present ``horizon''.  For compact topologies, the two
main effects on the CMB are: [1] the breaking of statistical isotropy
in characteristic patterns determined by the photon geodesic structure
of the manifold and [2] an infrared cutoff in the power spectrum of
perturbations imposed by the finite spatial extent.  To make a
detailed confrontation of these effects with the COBE maps requires
the computation of the pixel-pixel temperature correlation function
for each topology and for each orientation of it relative to the
sky. We present a general technique using the method of images for
doing this in compact hyperbolic (CH) topologies which does not
require spatial eigenmode decomposition.  We demonstrate that strong
constraints on compactness follow from [2] and that these limits can
be improved by exploiting the details of the geodesic structure for
each individual topology ([1]), as we show for the flat 3-torus and
selected CH models.}

Flat or open FRW models adequately describe the observed average
properties of our Universe. Much recent astrophysical data suggest the
cosmological density parameter $\Omega_0$ is $ < 1$.~\cite{opencase}
In the absence of a cosmological constant, this would imply a
hyperbolic 3-geometry for the universe. There are numerous theoretical
reasons, however, to favour compact topologies (reviewed in ref.~2).
To reconcile this with a flat or hyperbolic geometry, compact models
can be constructed by identifying points on the standard infinite FRW
space by the action of certain allowed discrete subgroups of
isometries, $\Gamma$. The FRW spatial hypersurface is then the
``universal cover'', tiled by copies of the compact
space.\footnote{Any point ${{\bf x}}$ of the compact space has an
image ${{\bf x}}_i = g_i {{\bf x}}$ in each ``Dirichlet'' domain on
the universal cover, where $g_i \in \Gamma$. Compact hyperbolic
manifolds (CHM) are described by discrete subgroups of the proper
Lorentz group. A census of CHMs and software (SnapPea) for computing
the generators of $\Gamma$ for any CHM is freely available.\cite{Minn}
CHMs can be classified in terms of $V/d_c^3$, where $V$ is the volume
and $d_c = H_0^{-1}/\sqrt{1-\Omega_0}$ is the curvature
radius.\cite{Thur7984}} Dynamical chaos arising from the resulting
complex geodesic structure in CH spaces has been proposed as an
explanation of the observed homogeneity of the universe.\cite{cor9596}

Any unperturbed FRW spacetime will have an isotropic cosmic microwave
background (CMB) regardless of global topological structure.  However,
the topology does affect the observed CMB temperature fluctuations
$\Delta T/T $ and thus can be tested.  At large angular scales,
$\Delta T/T $ is dominated by the Sachs-Wolfe effect: $\Delta T/T
\propto \Phi$, where $\Phi$ is the gravitational potential,
appropriately smoothed to take into account the COBE beam.  It is
usually computed by decomposing into eigenmodes of the 3-Laplacian for
the space, {\it e.g.} just plane waves for flat universes. We avoid
the nontrivial task of finding these eigenmodes for a CH space by
using the fact that the cor\-relation function $C_c \equiv
\langle\Phi({\bf x}) \Phi({\bf x}^\prime)\rangle$ in a compact space
can be expressed as a sum over the correlation function $C_u$ in its
universal covering space between the images of ${\bf x}$ and ${\bf
x}^\prime$: 
\be 
C_c({\bf x}, {\bf x}^\prime) = \lim _{N\to\infty} {1
\over N} \sum _{i=0}^N \sum _{j=0}^N C_u(g_i {\bf x}, g_j {\bf
x}^\prime)\, . 
\ee
The $g_i$ are ordered in increasing displacement and $g_0$ is
the identity. This procedure can be applied to any compact space
provided the set of elements $\{g_i\}$ of the symmetry group is
known. For a flat gravitational potential perturbation spectrum, 
$C_u({\bf x},{\bf x}^\prime) \propto \int d \ln(\beta ^2 + 1) \left[
{\rm sin(\beta r)/(\beta sinh }r) \right] $, where $r$ is the proper
separation between ${\bf x}$ and ${\bf x}^\prime$.

\begin{figure}[ht]
\centerline{
\epsfxsize=2.5in\epsfbox{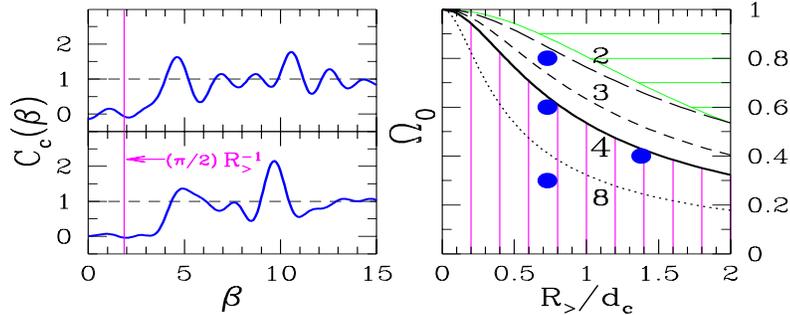}
}
\vspace{-0.98in}
\caption[]{{\bf {left panel:}} Examples of $C_c(\beta,{\bf {x}}, {\bf
{x}}^\prime)$ for zero pixel separation in a CH model with
$R_>=0.83d_c$.  Normalization is such that $C_u(\beta)=1$. The
infrared cutoff is present for ${\bf {x}} \neq{\bf {x}}^\prime$ as
well. {\bf {right panel:}} $\Omega_0$--$R_{>}$ constraints on CH
models.  Models in the region below the labeled lines have no power at
multipoles $\ell \le 2,3,4,8$, respectively. In the upper shaded
region, the compact space's volume exceeds the volume of a sphere of
radius ${\cal R}_H$. Models in the lower shaded region fail our cut
criterion with $\ell$=4. The large dots denote some examples for which
a full statistical comparison with the DMR data has shown the
topologies to be incompatible. This indicates the breaking of isotropy
can lead to models outside of the lower shaded region being ruled out.
\hfill\mbox{}}
\label{fig:OMro}
\end{figure}

The prescription (1) applies separately to each ``spectral mode",
$C_c(\beta,{\bf x}, {\bf x}^\prime)$, defined as the integrand in   
$C_c({\bf x}, {\bf x}^\prime) \propto \int d \ln(\beta ^2 + 1)
 C_c(\beta,{\bf x}, {\bf x}^\prime) $.
This decomposition is useful since the contribution to the $\ell^{\rm
th}$ multipole in the CMB angular correlation function comes
predominantly from scales $\beta _{\ell} \approx \ell d_c /{\cal R}_H
$ where ${\cal R}_H$ is the ``angle-diameter distance'' to the last
scattering surface. In all the CH models that we have studied,
compactness leads to a cutoff in $C_c(\beta,{\bf x}, {\bf x}^\prime)$
at scales at least four times the circum-radius of the space, $R_>$,
{\it i.e.}, for $\beta < \beta _{cut} \approx (\pi /2) R_>^{-1}$ (see
Fig.~1).  Thus the $\ell^{\rm th}$ multipole will be strongly
suppressed if $\beta_{cut} > \beta _{\ell} $. We consider a model to
contradict the COBE data if the $\ell \le 4$ multipoles are strongly
suppressed, translating to a criterion that the two parameters of the
problem, $R_>/d_c$ and ${\cal R}_H / d_c $, a function only of
$\Omega_0$, have to satisfy: $R_> > ( \pi/8 ){\cal R}_H $.
Fig.~\ref{fig:OMro} shows our constraints in the $\Omega_0 - R_{>}$
parameter plane.  We have studied some representative CH models (dots
in Fig.\ref{fig:OMro}) in detail by confronting the full predicted
statistics of the CMB anisotropy pattern with the all-channel COBE
map.\cite{ourpaper} We find that the exact likelihood falls steeply
once the cutoff wavelength is reduced to a size comparable to the
horizon, and the disallowed region in Fig.~\ref{fig:OMro} determined
by the simple $\beta_{cut}$ criterion is strongly ruled out. 

We conclude from Fig.~\ref{fig:OMro} that the bounding scale $R_>$ of
the universe cannot be much smaller than ${\cal R}_H$. This makes
problematical topological explanations of galaxy and quasar
distributions,\cite{lac_lum95,qsocct} but there may be cases for which
$R_{>}$ is large yet some closed geodesics are much shorter. Full
statistical analysis using the COBE maps tightens the limits in all
spaces considered so far.\cite{ourpaper} When we apply our full
statistical method to flat (equal-sided) 3-tori, we improve upon
previous limits~\cite{oliv95} on the size of the torus, $d_T \gta 4
H_0^{-1}$ (95\% CL), or $R_{>} \gta \sqrt{3} {\cal R}_H$, much
stronger than the conservative $\beta_{cut}$ criterion given above
because of the powerful breaking of statistical isotropy in the torus
case.  For compact hyperbolic universes, the low values of $\Omega_0$
suggested by various astrophysical observations, $ \lta 0.4 $, are
excluded for all the known~\cite{Minn} spaces. However, for $\Omega_0
\approx 1$, the multifaceted topology makes CH models more compatible
with the COBE data than flat $\Omega_0 \equiv 1$ torus models.


\small
\section*{References}
\begin{thebibliography}{99}
\bibitem{opencase} Dekel, A., Burnstein, D., \& White, S.D.M., 1996,
in {\it Critical Dialogues in Cosmology}, ed. N.Turok, World Scientific.
\bibitem{lac_lum95}Lachieze-Rey, M. \& Luminet, J.-P. 1995,
Phys. Rep. {\bf 25}, 136
\bibitem{cor9596} Cornish, N.J., Spergel, D.N. \& Starkman, G.D. 1996,
Phys.Rev.Lett. {\bf 77}, 215
\bibitem{Minn} Weeks, J.R., {\sl  SnapPea: A computer program for creating and
studying hyperbolic 3-manifolds}, University of Minnesota Geometry Center.
\bibitem{Thur7984}Thurston, W.P. 1979, {\sl The Geometry of
3-Manifolds}, lecture notes, Princeton University; Thurston, W.P. \&
Weeks, J.R. 1984, Sci. Am. (July), 108
\bibitem{ourpaper} Bond, J.R., Pogosyan, D. \& Souradeep, T. 1997, preprint
\bibitem{oliv95}Sokolov, I.Y. 1993, JETP Lett. {\bf 57}, 617;
Starobinsky, A.A. 1993, JETP Lett. {\bf 57}, 622; Stevens, D., Scott,
D. \& Silk, J. 1993, Phys. Rev. Lett. {\bf 71}, 20; de Oliveira Costa,
A. \& Smoot, G.F. 1995, Ap.J. {\bf 448}, 477; de Oliveira
Costa, A., Smoot, G.F. \& Starobinsky, A.A. 1996, Ap.J. {\bf
468}, 457
\bibitem{qsocct} 
{\it e.g.},  Fagundes, H.V. 1989, Ap.J. {\bf 338},
618; Roukema, B.F. 1996, preprint
\end{thebibliography}
\end{document}